\documentclass[
    ,final            % use final for the camera ready runs
  ]
  {aipproc}

\layoutstyle{6x9}

\begin{document}

\title{Heavily obscured AGN in the local Universe}

\classification{98.54.Cm: 95.85.Nv, 95.85.Hp}
\keywords{Active Galaxies: X-ray, Infrared}

\author{P. Severgnini}{
  address={INAF-Osservatorio Astronomico di Brera}
 ,altaddress={e-mail: paola.severgnini@brera.inaf.it} % additional visiting address
}

\author{A. Caccianiga}{
  address={INAF-Osservatorio Astronomico di Brera}
}

\author{R. Della Ceca}{
  address={INAF-Osservatorio Astronomico di Brera}
}

\author{A. Moretti}{
  address={INAF-Osservatorio Astronomico di Brera}
 }

\author{C. Vignali}{
  address={Dip. di Astronomia, Univ. degli Studi di Bologna}
}

\author{V. La Parola}{
 address={INAF-IASF Palermo}
}

\author{G. Cusumano}{
  address={INAF-IASF Palermo}
}

\begin{abstract} 

We present here a new powerful diagnostic plot to select heavily obscured AGN in
the local universe by combining infrared (Spitzer, IRAS) and X-ray (XMM)
information. On the basis of this plot, we selected a sample of X--ray obscured
sources in the 2XMM catalogue and found seven newly discovered Compton-thick AGN
candidates.

\end{abstract}

\maketitle

%%%%%%%%%%%%%%%%%%%%%%%%%%%%%%%%%%%%%%%%%%%%
%% MAINMATTER
%%%%%%%%%%%%%%%%%%%%%%%%%%%%%%%%%%%%%%%%%%%%

\section{The diagnostic plot and the IRAS-2XMM Sample} 
\paragraph{\bf The Diagnostic Plot} Despite their cosmological relevance, less than two dozens of
confirmed heavily obscured, Compton--thick  (Nh$>$10$^{24}$ cm$^{-2}$) AGN have been found
so far. We propose a new diagnostic plot to select Compton-thick AGN candidates  in the
local universe. The plot is based on the combination of the F(2--10 keV)/($\nu_{IR}$
F$_{IR}$) flux ratio (where IR is either 24 $\mu$m (Spitzer) or 25 $\mu$m (IRAS)) with the
XMM-Newton color (hardness ratio, HR). While the F(2-10 keV)/($\nu_{IR}$ F$_{IR}$) flux
ratio allows us to separate heavily obscured AGN candidates and starburst galaxies from
less obscured (Nh$<$5$\times$10$^{23}$ cm$^{-2}$) AGN (\cite{Severgnini2007}), the X-ray
color is able to separate starburst (HR$<$-0.1) from obscured AGN (HR$>$-0.1). The region
of Compton-thick AGN candidates is marked in Figure 1 (left panel).

\paragraph{\bf The IRAS-2XMM Sample} To test the proposed diagnostic plot in selecting
Compton-thick AGN, we have cross-correlated the IRAS Point Source Catalog (PSC) v2.1 with
the bright end (F(4.5-12 keV)$>$10$^{-13}$ erg cm$^{-2}$ s$^{-1}$)  of the incremental
version of the 2XMM catalogue (\cite{Watson2009}). We find that 47 IRAS sources populate
the region of the Compton-thick AGN candidates, 46 of which are extra-galactic sources
(hereafter the IRAS-2XMM sample, Fig. 1, right panel). For all but one source, the
spectroscopic classification is already available in the literature. Up to now, we have
performed a preliminary X--ray analysis for 40 out of the 46  sources.  Twenty--four of
them are already known as Compton--thick AGN in the literature (10 have also high energy
data that confirm their nature). Among the remaining 16 sources, there are 9   obscured but
Compton-thin sources and  7 newly discovered Compton-thick candidates (i.e. with 2--10 keV 
properties consistent with a typical Compton--thick emission). For one of them a
preliminary analysis of SWIFT-BAT (\cite{Cusumano2009}) and Suzaku (proprietary data) data 
confirms our classification as a Compton--thick AGN (Severgnini et al. in prep.).  In
conclusion, of the 40 sources already analyzed so far  we find that all of them  are
obscured and at least $\sim$75\% of them  are Compton--thick AGN. We thus demonstrate that 
the X-ray/IR flux ratio, along with the X--ray color, can be used to efficiently select
local (z<0.1), heavily obscured and Compton-thick AGN.

%%%%%%%%%%%%%%%%%%%%%%%%%%%%%%%%%%%%%%%%%%%%
%% Sample figure:
%%
%% The option [height=...] scales the picture to the given height,
%% without it it would be printed at its nominal size
%%%%%%%%%%%%%%%%%%%%%%%%%%%%%%%%%%%%%%%%%%%%

\begin{figure}
  \includegraphics[height=.34\textheight]{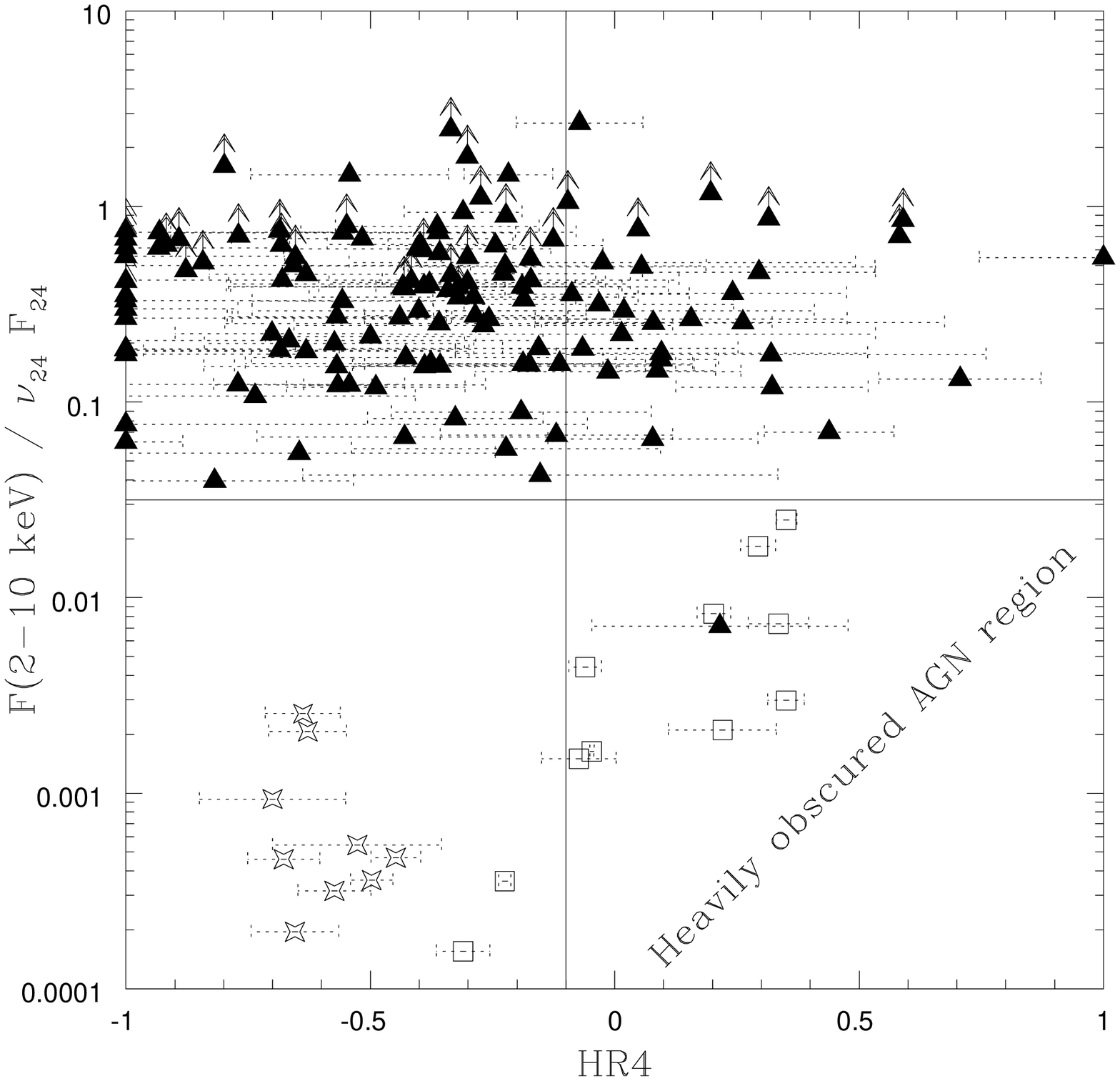}
  \includegraphics[height=.34\textheight]{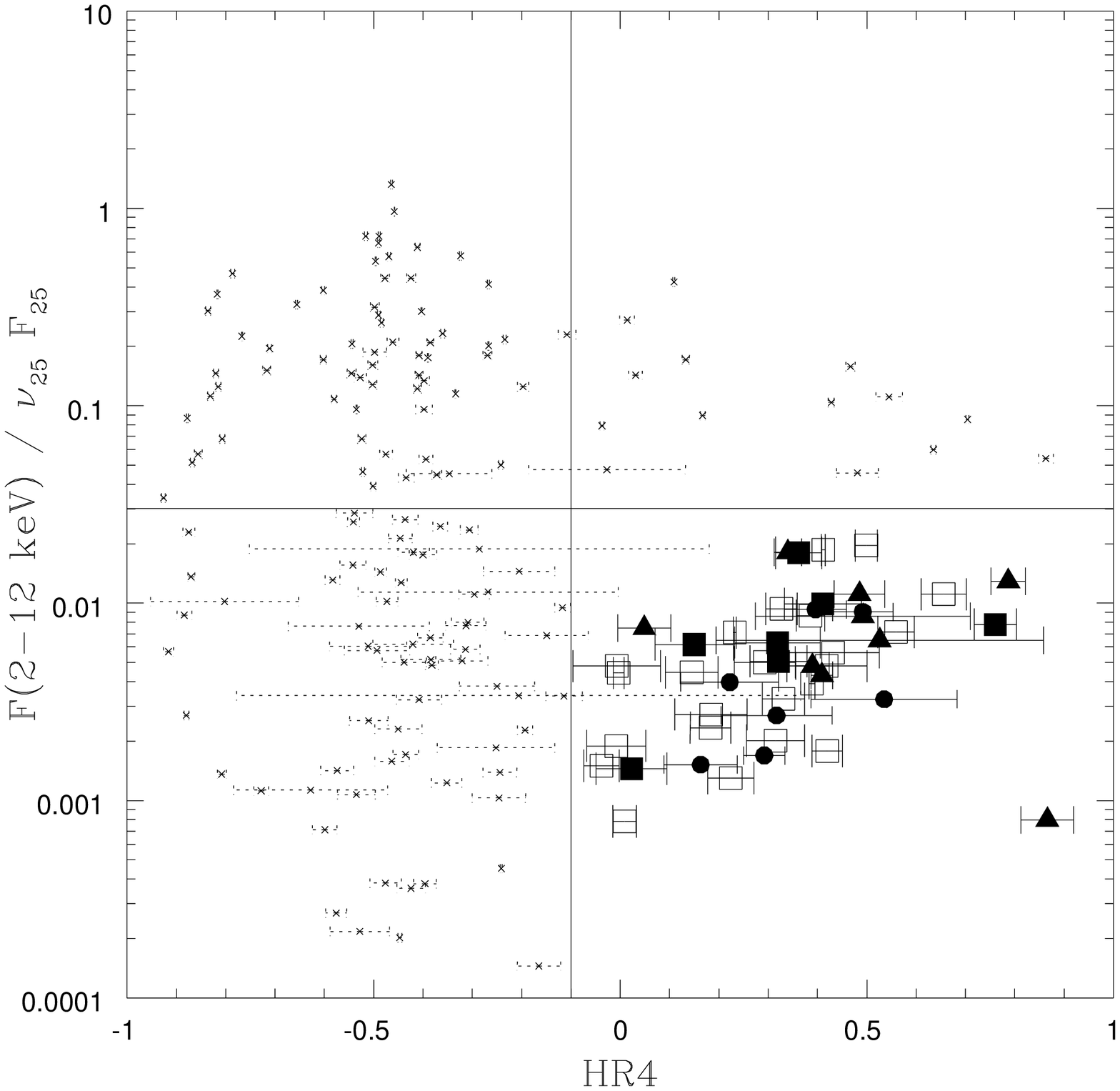}

  \caption{{\it Left Panel:} F(2-10 keV)/($\nu_{24}$ F$_{24}$) vs. HR4 (defined
using the two following bands: 2-4.5 keV and 4.5-12 keV) for a large sample of
AGN already studied in the literature. Filled triangles in the upper quadrants
are Compton-thin AGN (Nh$<<$10$^{24}$ cm$^{-2}$) taken from the HBS sample
(\cite{DellaCeca2008a}) and the XMDS survey (\cite{Polletta2007}). The only
triangle in the lower--right quadrant is a Compton-thick candidate according to
\cite{Polletta2007}. Stars are a sample of local starburst galaxies
(\cite{Ranalli2003}) and open squares are local "confirmed" Compton-thick AGN
from \cite{DellaCeca2008b}. {\it Right panel:} F(2-12 keV)/($\nu_{25}$F$_{25}$)
vs. HR4 diagnostic plot for the sources resulting from the IRAS-2XMM
cross-correlation.  The sources belonging to the IRAS-2XMM sample populate the 
lower--right quadrant. Open squares are the 24 Compton--thick already known in
the literature, filled squares are the 7 newly discovered Compton--thick
candidates and the filled triangles are the 9 Compton-thin AGN. No X--ray
analysis has been performed yet for small filled circles.}

\end{figure}

%%%%%%%%%%%%%%%%%%%%%%%%%%%%%%%%%%%%%%%%%%%%%%%%
%% BACKMATTER
%%%%%%%%%%%%%%%%%%%%%%%%%%%%%%%%%%%%%%%%%%%%%%%%

\begin{theacknowledgments}
We acknowledge financial support from ASI (grant n.
I/088/06/0 and COFIS contract).
\end{theacknowledgments}



\bibliographystyle{aipproc}   % if natbib is available

%\bibliography{sample}

%  \typeout{******************************************}
%  \typeout{** Please run "bibtex \jobname" to optain}
%  \typeout{** the bibliography and then re-run LaTeX}
%  \typeout{** twice to fix the references!}
%  \typeout{******************************************}

%%%%%%%%%%%%%%%%%%%%%%%%%%%%%%%%%%%%%%%%%%%
%% The following lines show an example how to produce a bibliography
%% without the help of the BibTeX program. This could be used instead
%% of the above.
%%%%%%%%%%%%%%%%%%%%%%%%%%%%%%%%%%%%%%%%%%%

\end{document}